\documentclass[
reprint,
longbibliography,
 showpacs,
 amsmath,amssymb,
 aps,
 prl,
floatfix,
]{revtex4-2}

\usepackage{amsmath,amssymb}
\usepackage{physics}
\usepackage{graphicx, color}
\usepackage{dcolumn}
\usepackage{bm}
\usepackage{bbm}
\usepackage{hyperref}
\usepackage[dvipsnames]{xcolor}
\hypersetup{
setpagesize=false,
 bookmarksnumbered=true,%
 bookmarksopen=true,%
 colorlinks=true,%
 linkcolor=Blue,
 citecolor=Blue,
 urlcolor = Blue,
}
\usepackage{pdfpages} 
\usepackage{pgffor} 

\makeatletter
\AtBeginDocument{\let\LS@rot\@undefined}
\makeatother

\def\supplementfilename{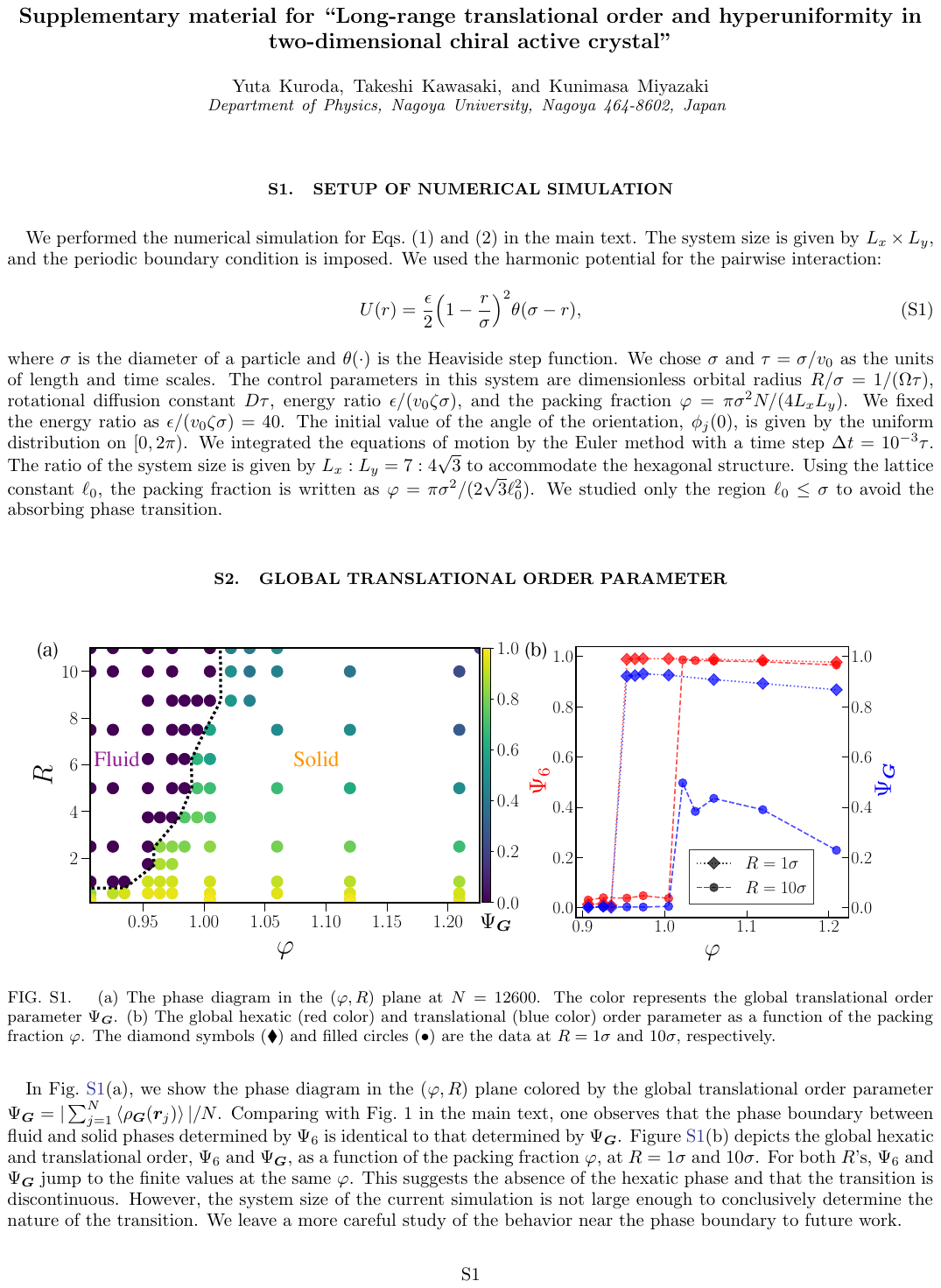}

\pdfximage{\supplementfilename}
\def\numbersupplementpages{\the\pdflastximagepages}

\newif\ifarXiv
\arXivtrue 

\newcommand{\RM}[1]{\mathrm{#1}}

\newcommand{\eq}[1]{Eq.~(\ref{#1})}
\newcommand{\fig}[1]{Fig.~\ref{#1}}
\newcommand{\figs}[1]{Figs.~\ref{#1}}
\newcommand{\p}{\partial}

\begin{document}

\preprint{APS/123-QED}

\title{Long-range translational order and hyperuniformity \\ in two-dimensional chiral active crystal}

\author{Yuta Kuroda}
\email{kuroda@r.phys.nagoya-u.ac.jp}

\author{Takeshi Kawasaki}%

\author{Kunimasa Miyazaki}%
 
\affiliation{%
 Department of Physics, Nagoya University, Nagoya 464-8602, Japan
}%

\date{\today}

\begin{abstract}
We numerically study two-dimensional athermal chiral active particles at high densities. 
The particles in this system perform the circular motion with frequency $\Omega$. 
We show that the system crystallizes at high densities even in two dimensions, accompanied by the true long-range translational order.  
This is due to the anomalous suppression of displacement fluctuations associated with hyperuniformity. 
These findings can be explained using an active elastic theory quantitatively.
Surprisingly, the crystals become unstable and melt in the limit of $\Omega=0$, for the spatial dimension of four or less. 
This result can be explained by a mechanism akin to quenched random systems for which the lower critical dimension is four.  
\end{abstract}

\maketitle

{\it Introduction.}---
The study of active matter is now one of the main streams of soft matter and nonequilibrium physics~\cite{Ramaswamy2010Anuual_Review, Marchetti2013RMP, Cates2015Annual_Review, Bechinger2016RMP,zottl2016, Doostmohammadi:2018aa, Markus2020anual, Shankar:2022aa, BowickPRX2022, Andreas2023annual}.
In particular, active matter at very high densities, or {\it active solids}, has recently attracted much attention. 
One of the primary goals of the study of active solids is to understand how the active driving forces affect the well-known properties of ordered or disordered solids in equilibrium.  
It is natural to expect that the fate of the translational order in the crystals~\cite{Bialke2012PRL, Ferrante_2013, Weber2014PRL, Briand2018PRL, Klamser:2018ab, Maitra2019PRL, Caprini2020PRR, Digregorio2022SoftMatter}, the dynamical heterogeneities in glasses~\cite{Berthier:2013aa, Ni:2013aa, Berthier2014PRL, Szamel2015PRE, Mandal2016SoftMatter, Klongvessa2019PRL, Janssen_2019, Berthier2019, Klongvessa2022, Lama2024}, the rheological properties of amorphous solids, or even the criticality of the jamming transition~\cite{Henkes2011PRE, Liao2018SoftMatter, Mo2020SoftMatter, YangPRE2022, Anand2024PRR} are dramatically altered by the nonequilibrium fluctuations. 
In the study of active solids, simple particle models, such as active Brownian particles (ABP)~\cite{Howse2007PRL, Hagen:2011aa, fily2012PRL} or active Ornstein--Uhlenbeck particles, have been often used~\cite{Fodor2016PRL}. 
In these models, the strength of the self-propelled force and its persistence time are the two parameters that characterize how far the system is from equilibrium~\cite{fily2012PRL, Fodor2016PRL}. 
Previous studies have reported that active crystals and active glasses behave like systems in equilibrium with additional forces or equilibrium systems under external perturbations. 
For example, the crystalline and glassy solids tend to melt as the activity increases~\cite{Fily2014soft_matter, Digregorio2018PRL, Caprini2020PRR}. 
The flowing or yielding properties of active solids are similar to those of equilibrium solids deformed by the nonequilibrium external force such as the shear stress~\cite{Mo2020SoftMatter, Mandal_2021, Villarroel2021SoftMatter, Keta2023SoftMatter}.  
Furthermore, even the melting of the two-dimensional active crystals is also found to follow qualitatively the same two-step melting scenario as the equilibrium solids~\cite{Digregorio2018PRL, Shi2023PRL}, such as the KTHNY theory~\cite{Halperin1978, Nelson1979, Young1979}.  
Given that many studies have shown similarities between active and equilibrium solids, it is natural to ask where there is an active solid that has completely different properties and for which the conventional theories of equilibrium solids do not apply. 

In this paper, we show that the two-dimensional crystallization of extremely persistent ABP with and without chirality is qualitatively different from that of any equilibrium system. 
The system we consider is the monodisperse chiral ABP (cABP). 
Each particle is driven by a chiral driving force so that it can perform a unidirectional circular motion, which violates the mirror symmetry~\cite{Teeffelen2008PRE, Ma2017, Li2019, Ma2022}.   
The equation of motion of the $j$th particle is given by
\begin{equation}
 \dot{\bm r}_j(t) =\frac{1}{\zeta} {\bm F}_j(t) + v_0 \bm e(\phi_j(t)), \label{1}
\end{equation}
where $\zeta$ is the drag coefficient and ${\bm F}_j$ is the force acting on the particle. 
$v_0{\bm e}(\phi)= v_0(\cos\phi,\sin\phi)$ is the self-propelled velocity. 
$v_0$ is fixed as a constant and the orientation ${\bm e}(\phi_j)$ obeys
\begin{equation}
\dot{\phi}_j(t)= \Omega + \sqrt{2D}\eta_j(t), \label{2}
\end{equation}
where $\eta_j$ is a Gaussian white noise and $D=1/\tau_\RM{p}$ is the inverse of the persistence time. 
In particular, we consider the infinite limit of the persistence time $D=0$ or $\tau_\RM{p}\rightarrow\infty$.
$\Omega$ represents the frequency at which the particle performs a circular motion.
It is known that cABP with $D=0$ exhibit hyperuniformity (HU)~\cite{Li2019, kuroda2023_chiral}: the static structure factor vanishes at small wavevectors $\bm q$ as $S(\bm q\rightarrow \bm 0)\sim q^\alpha\ (\alpha>0)$~\cite{TORQUATO}.

Our main findings are twofold. 
First, the system crystallizes with the long-range translational order at high densities even in two dimensions whenever $\Omega$ is finite.
This is in contrast to two-dimensional equilibrium crystals, which have no long-range translational order according to the Hohenberg--Mermin--Wagner theorem~\cite{Hohenberg1967, Mermin1966, Mermin1968}.
The crystallization of the two-dimensional cABP is due to the strong suppression of displacement fluctuations by the same mechanism of HU.
This crystallization mechanism is similar to that of the center-of-mass conserved random organization (cRO) model~\cite{Galliano2023PRL}, which also exhibits HU~\cite{Hexner2017, Lei2019_pnas, lei2023does}.
Continuous symmetry breaking in low-dimensional nonequilibrium systems has attracted much attention in recent years~\cite{Vicsek1995PRL, TonerPRL1995, Dadhichi2020PRE, Nishiguchi2023, Nakano2021PRL, minami2022, ikeda2024, Bassler1995PRE, Matthew2010PRE, Loos2023PRL, Giomi2022PRL, maire2024}. 
The authors of Ref.~\cite{Galliano2023PRL} argued that HU plays a key role in the emergence of the long-range order. 
This argument has been bolstered by recent theoretical studies~\cite{ikeda2023correlated, Ikeda2024PRE, ikeda2023}. 
Second, we find that when $\Omega=0$, the crystalline phase becomes even more unstable than its equilibrium counterpart. 
Although the numerical simulation is only performed in two dimensions, our argument suggests that extremely persistent ABP cannot crystallize in four dimensions or less. 
The forbidden long-range translational order of the system with $\Omega =0 $ can be understood by the similarity to quenched random systems with continuous symmetry, where the lower critical dimension is known to be four, according to the Imry--Ma argument~\cite{Imry-Ma1975}.
For systems in both the presence and absence of $\Omega$, we also develop a linear elastic theory for cABP that quantitatively explains the numerical results.

Recently, several studies on the chiral active matter have reported collective behavior ~\cite{Lowen2016, Liebchen_2022}, such as the formation of vortex patterns~\cite{Ingmar2005Science, Liebchen2017, Zhang:2020ux, Kruk2020PRE, Ventejou2021PRL, Liao2021SoftMatter} and anomalous transport phenomena~\cite{Banerjee2017, Hargus2021PRL, Tan:2022aa, Fruchart2023,
Siebers2024PNAS} (see also, {\it e.g.}, Refs.~\cite{Leonardo2011, Kummel2013PRL, Mano2017, Yamamoto2017SoftMatter, Patra:2022vd, DiLuzio2005, Leonardo2011, Siebers2023, Chan:2024aa} for experiments on chiral active matter). 
At high densities, odd elasticity~\cite{Tan:2022aa}, glassy dynamics~\cite{Debets2023PRL}, and spatial velocity correlations~\cite{shee2024}  have also been studied. 
However, the nature of the crystallization in two dimensions has not been carefully studied so far, either with or without chirality. 

\begin{figure}[t]
\centering
\hspace*{-3.4mm}
  \includegraphics[width=8.5cm]{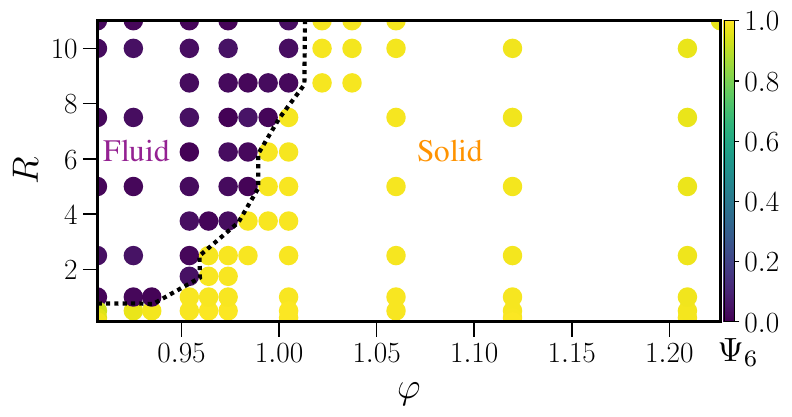}
  \caption{\label{fig1}The phase diagram in the $(\varphi,R)$ plane.
  The color represents the modulus of the global hexatic order parameter $\Psi_6$.
  The number of particles is $N=12600$.
  }
\end{figure}

{\it Numerical results.}---
We study the two-dimensional cABP obeying Eqs.~(\ref{1}) and (\ref{2})
with $D=0$.
The force is given by ${\bm F}_j = - \sum_{k(\neq j)}\nabla_j U(|\bm r_j-\bm r_k|)$, where $U(r)$ is the pairwise potential. 
In this study, we use the harmonic potential: $U(r) = \epsilon(1-r/\sigma)^2/2$ for $r < \sigma $ and $U(r) = 0$ for $ r\geq\sigma$, where $\sigma$ is the diameter of a particle. 
In the numerical simulation, we choose $\sigma$ and $\tau = \sigma/v_0$ as the unit of length and time scales. 
The parameters in this system are 
the orbital radius $R=v_0/\Omega$, the energy ratio $\epsilon/(v_0\zeta\sigma)$, and the packing fraction $\varphi = \pi\sigma^2N/(4L_xL_y)$, where $N$ is the total number of particles and $L_x$ and $L_y$ are the side lengths of the system. 
Further details of the simulation setup are described in Sec.~S1 of the Supplementary Material (SM)~\cite{SI}.

\begin{figure}[t]
\centering
\hspace*{-1.4mm}
  \includegraphics[width=9.1cm]{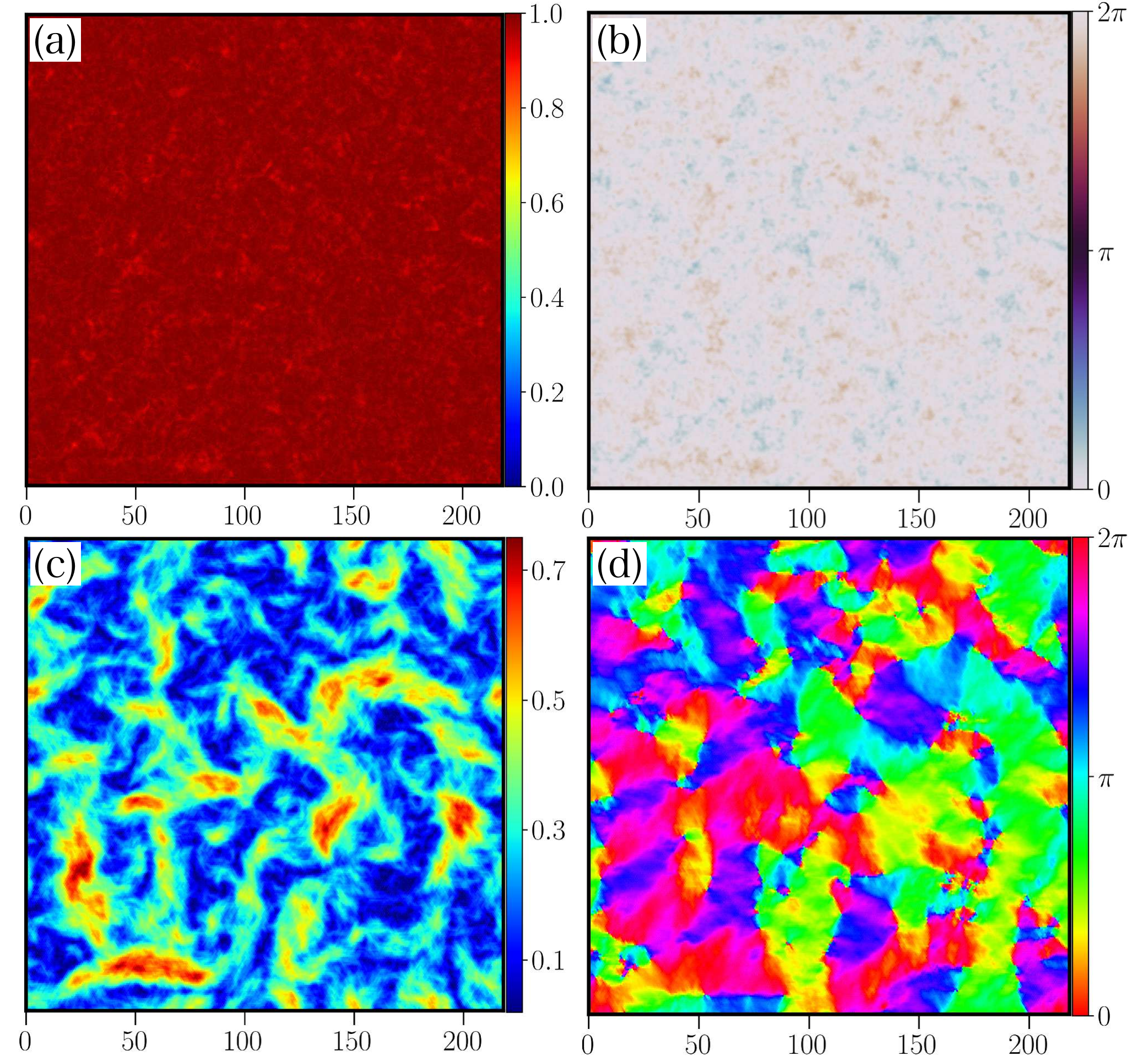}
  \caption{\label{fig2}Snapshots of the particle configurations at $R=10\sigma$, $\varphi=1.209$, and 
$N=72576$. 
  (a)~Modulus of the hexatic order parameter $|\psi_6(\bm r_j)|$.
  (b)~Argument of the hexatic order parameter $\arg\psi_6(\bm r_j)$.
  (c)~Modulus of the displacement $|\bm u_j|$.
  (d)~Angle of the displacement $\bm u_j$ with respect to the $x$axis. 
}
\end{figure}

\begin{figure}[t]
\centering
\hspace*{-3.4mm}
  \includegraphics[width=8.4cm]{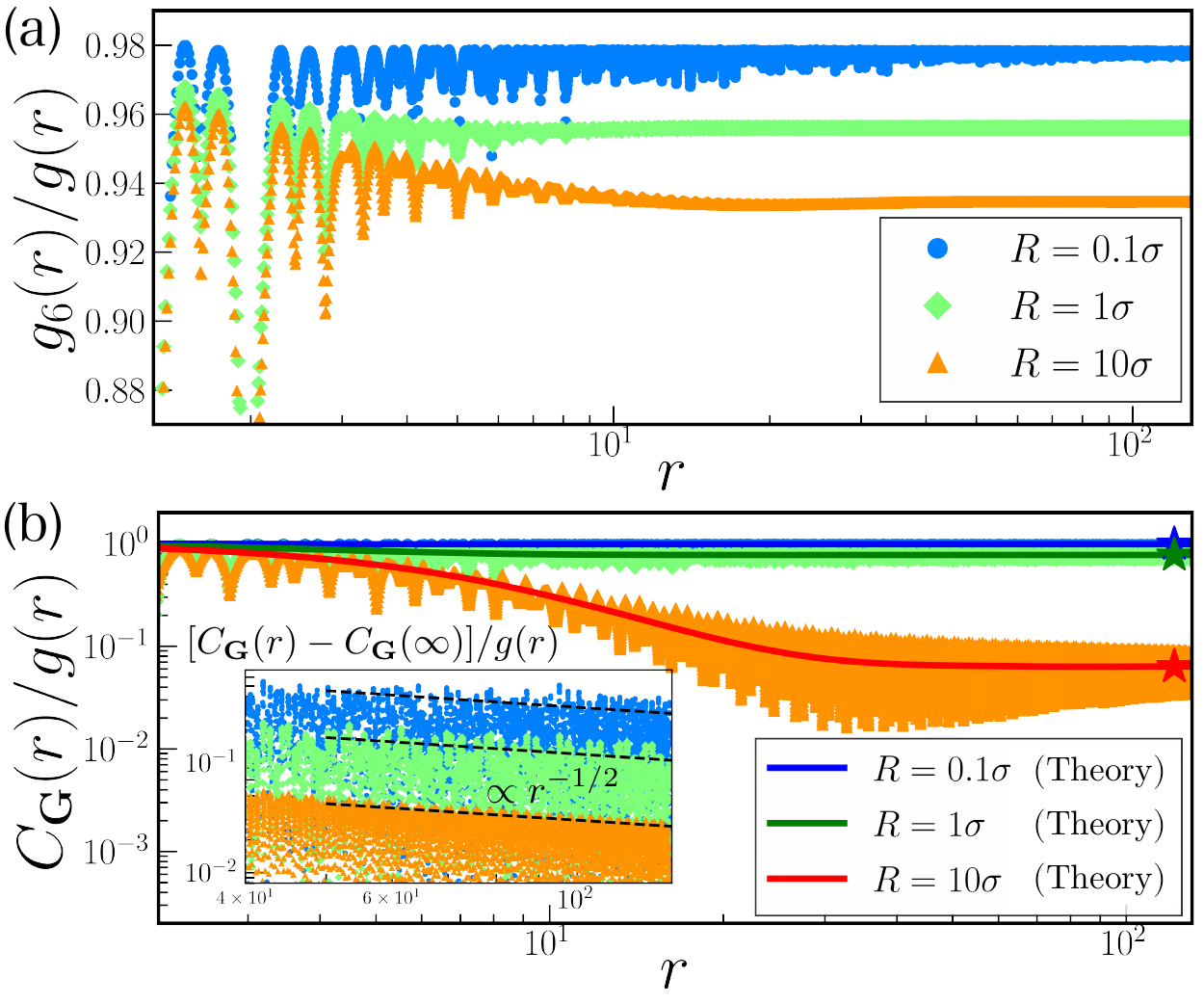}
  \caption{\label{fig3}Correlation functions of the hexatic order parameter~(a) and the translational order parameter~(b) for different $R$ at $\varphi=1.209$ and 
$N=103544$.
  $r$ denotes the distance.
  Both correlation functions are normalized by the radial distribution function $g(r)$.
  In panel~(b), the solid lines are the theoretical predictions and the star symbols represent the asymptotic values $C_{\bm G}(r\rightarrow\infty)$, \eq{15}.
  The inset shows the magnification of $C_{\bm G}(r) - C_{\bm G}(\infty)$, where $C_{\bm G}(\infty)$ comes from the theoretical expression, \eq{15}. The dashed lines are proportional to $r^{-1/2}$.
}
\end{figure}

The first task is to identify the solid phase.
Figure 1 shows the phase diagram in the $(\varphi, R)$ plane in
the high density regime.
The color represents the modulus of the global hexatic order parameter defined by $\Psi_6 = |\sum_{j}\expval{\psi_6(\bm r_j)}/N|$, where
$\psi_6(\bm r_j) = \sum_{k\in\mathcal N_6(j)}e^{6i\theta_{jk}} / 6$ is the local hexatic order parameter~\cite{Steinhardt1983, Engel2013PRE}.
$\mathcal N_6(j)$ is the set of the six nearest neighbors of the $j$th
particle, and $\theta_{jk}$ is the angle of the vector $\bm r_k-\bm r_j$ with respect to the $x$-axis.
The transition appears to be discontinuous.
We also find that the global translational order parameter jumps to the finite values at the same $\varphi$ (see Sec.~S2 of the SM~\cite{SI}).
However, the current system size is not large enough to definitively determine the nature of the transition.
In this work, we do not pursue this point and focus on the nature of the translational order in the solid phase. 
In the following, the packing fraction is fixed at $\varphi = 1.209$ where the system is in the solid phase for all finite $R$ (our conclusion is not affected by densities, see Sec.~S3 of the SM~\cite{SI}).
In \figs{fig2}(a) and (b), we show the particle configuration colored by the modulus and argument of $\psi_6(\bm r_j)$. 
Obviously, the system is in the solid phase where the orientational order is long-ranged.
To ensure that the system is in the solid phase, we calculate the spatial correlation function of the hexatic order parameter defined by 
$g_6(r) = \expval*{\sum_{j\neq k} \psi_6(\bm r_j) \psi_6^*(\bm r_k)\delta(\bm r-\bm r_j+\bm r_k)}/(\rho N)$,
where $\rho =N/(L_xL_y)$ is the mean density and $*$ represents the complex conjugate.
We show $g_6(r)$ for different $R$ in \fig{fig3}(a).
For all $R$, we find $g_6(r \gg 1)\simeq \RM{const.}$, which is the characteristic behavior of the solid phase~\cite{Kapfer2015PRL}.

\begin{figure}[t]
\centering
\hspace*{-3.4mm}
  \includegraphics[width=9cm]{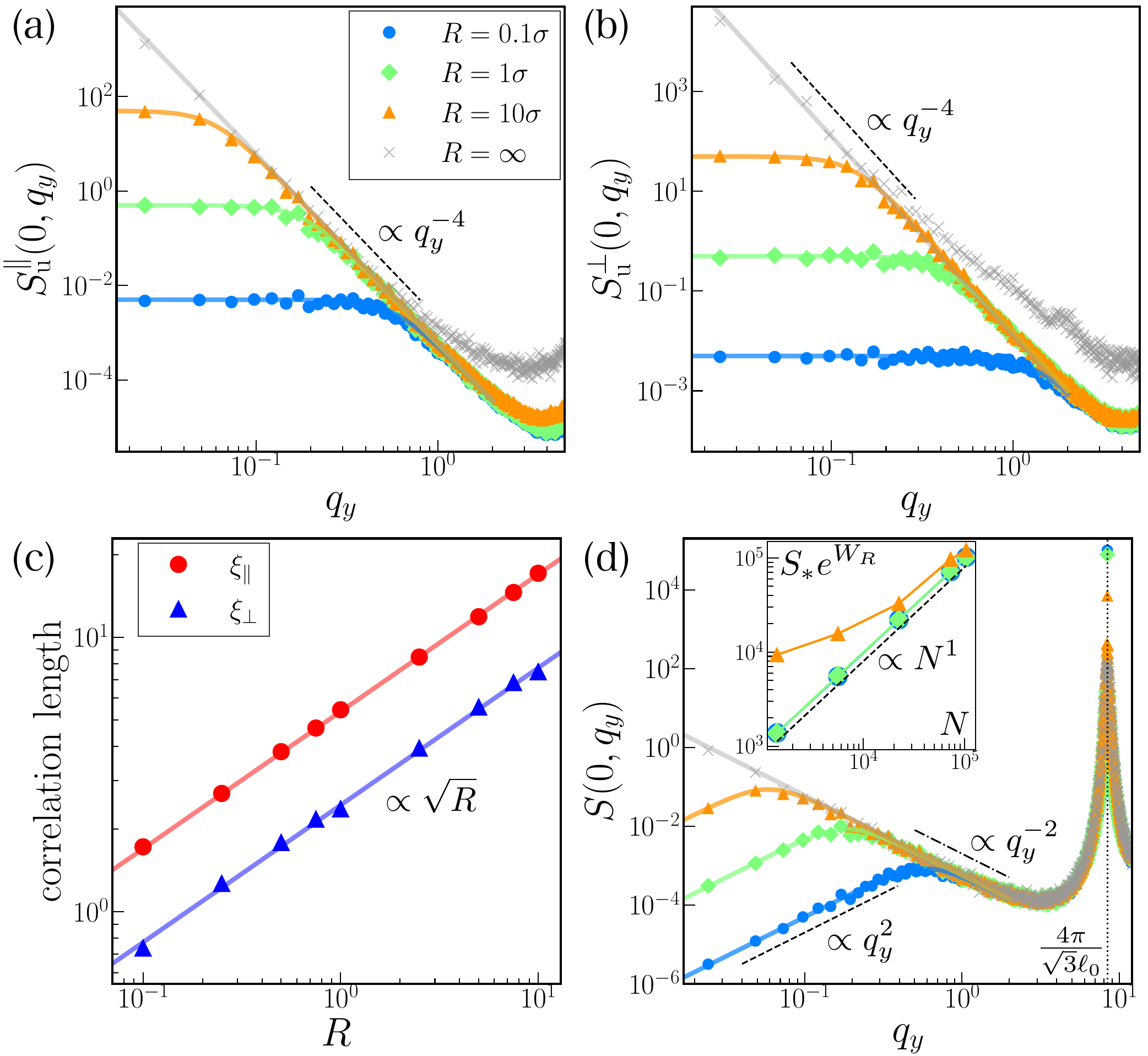}
  \caption{\label{fig4}(a)~The longitudinal and (b)~transverse displacement correlation function in Fourier space at $\varphi=1.209$ and $N=103544$.
  The solid symbols represent the numerical data, and the solid lines are the theoretical predictions, \eq{13}. 
  (c)~Correlation length of the displacement correlation functions.  The solid lines represent a fit by $\xi_{\parallel,\perp}\propto \sqrt{R}$.
  (d) Static structure factor.  The solid symbols represent the numerical data, and the solid lines are the theoretical predictions. 
  The vertical doted line represents $q_y=4\pi/(\sqrt{3}\ell_0)$.
  The inset depicts the $N$ dependence of the peak height, $S_*=S(\bm G)$, rescaled by the $R$ dependence, $e^{-W_R}$.
}
\end{figure}

Next, in \fig{fig3}(b), we plot the correlation function of the translational order parameter defined by
\begin{equation}
C_{\bm G}(r) = \frac{1}{\rho N}\expval{\sum_{j\neq k} \rho_{\bm G}(\bm r_j) \rho_{\bm G}^*(\bm r_k)\delta(\bm r-\bm r_j+\bm r_k)}. \label{4}
\end{equation}
Here $\rho_{\bm G}(\bm r_j) = e^{i\bm G\cdot \bm r_j}$ and $\bm G$ is the reciprocal lattice vector whose modulus is chosen to be $4\pi/(\sqrt{3}\ell_0)$ where $\ell_0$ is the lattice constant. 
For all $R$, $C_{\bm G}(r)$ converges to a constant for large $r$, ensuring the presence of the long-range translational order.
The inset of \fig{fig3}(b) shows $C_{\bm G}(r) - C_{\bm G}(\infty)$.
$C_{\bm G}(r)$ converges to $C_{\bm G}(\infty)$ according to the power-law function $r^{-1/2}$.
This property is different from that of equilibrium crystals~\cite{Mermin1968, Prestipino2011PRL} and active crystals without chirality~\cite{Digregorio2018PRL, Shi2023PRL} in two dimensions, where one finds only the quasi-long-range order characterized by $C_{\bm G}(\infty)=0$ and a power-law decay $C_{\bm G}(r)\sim r^{-\eta_G}$ with a non-universal exponent $\eta_G(>0)$~\cite{chaikin2000}.
 
The presence of long-range order can also be understood in terms of displacement correlation functions~\cite{Galliano2023PRL}. 
Let us denote the displacement in Fourier space as $\hat{\bm u}(\bm q) = \sum_j\bm u_j e^{-i\bm q\cdot \bm r_j}$ and decompose it into the longitudinal and transverse components as $\hat{\bm u}(\bm q) = \hat{u}_\parallel(\bm q) \bm e_\parallel  +  \hat{u}_\perp(\bm q) \bm e_\perp$.
Here $\bm e_\parallel $ and $\bm e_\perp$ are unit vectors parallel and perpendicular to the wave vector $\bm q$, respectively. 
$\bm u_j=\bm r_j-\bm r_j^{(0)}$ is the displacement of the $j$th particle, where $\bm r_j^{(0)}$ is the equilibrium position. 
The correlation functions of each component are defined by $S_\RM{u}^\RM{X}(\bm q) = \expval{|\hat{u}_\RM{X}(\bm q)|^2}/N,~\RM{X} \in \{\parallel,\perp\}$.
These quantities are related to the amplitude of the mean square displacement by
\begin{equation}
A_L = \lim_{t\rightarrow \infty}\frac{1}{N}\sum_{j=1}^{N}\expval{|\bm u_j(t)|^2} \propto \int_{2\pi/L}^{\Lambda_\RM{c}}\dd q\ qS_\RM{u}(\bm q), \label{5}
\end{equation}
where $S_\RM{u}(\bm q) = S_\RM{u}^\parallel(\bm q) + S_\RM{u}^\perp(\bm q)$ and $\Lambda_\RM{c}$ is the ultraviolet cutoff. 
For equilibrium systems at temperature $T$, the equipartition rule holds
for each component: $m\omega^2_\RM{X}(\bm q)\expval{|\hat{u}_\RM{X}(\bm
q)|^2}/2 = T/2$ with the linear dispersion relation $\omega_\RM{X}(\bm q)\propto q$.
Therefore, $S_\RM{u}^\RM{X}(\bm q) \propto 1/q^2\ (q\rightarrow 0)$ for all $T> 0$.
Then $A_L$ diverges as $\log L$ in the limit $L\rightarrow \infty$, implying that the crystalline order is unstable.
On the other hand, if $S_\RM{u}(\bm q)$ diverges more slowly than $1/q^2$ or converges to a constant as $q\rightarrow 0$, $A_L$ is constant in the large $L$ limit, and the order is stable.
We show $S_\RM{u}^\parallel(\bm q)$ and $S_\RM{u}^\perp(\bm q)$ of our system in \figs{fig4}(a) and (b), respectively. 
Both $S_\RM{u}^\parallel(\bm q)$ and $S_\RM{u}^\perp(\bm q)$ are constant in the limit $q\rightarrow 0$ for all $R$. 
This means that $A_L$ does not depend on the system size in the large-scale limit and the crystalline order is stable.  
Furthermore, $S_\RM{u}^\RM{X}(\bm q)$ behaves as $1/q^4$ at large $q$.
The onset of $1/q^4$ behavior implies the presence of the characteristic length scales.
The length scales correspond to the spatially correlated regions observed in real space as shown in \figs{fig2}(c) and~(d). 
To extract these length scales, we fit the data with the function
$S_\RM{u}^\RM{X}(\bm q) = R^2/[2(1+(\xi_\RM{X}q)^4)]$, which will be derived theoretically below.
The solid lines in \figs{fig4}(a) and (b) represent the fitting function.
The only fitting parameter is the correlation length $\xi_\RM{\parallel,\perp}$.
In \fig{fig4}(c), we plot $\xi_{\parallel,\perp}$ as a function of $R$.
Both correlation lengths grow as $\xi_{\parallel,\perp}\propto \sqrt{R}$.
Note that the displacement correlation behaves as $S_\RM{u}^\RM{X}(\bm q)\propto 1/q^4\ (q\rightarrow 0)$ in the limit $R\rightarrow\infty$ or $\Omega \rightarrow 0$ (the limit of ABP with infinite persistence time). 
The gray symbols in \figs{fig4}(a) and (b) represent the numerical data for $R=\infty$, confirming $S_\RM{u}^\RM{X}(\bm q)\propto 1/q^4$.
This implies that in the limit of $R\rightarrow\infty$ or $\Omega \rightarrow 0$, the crystalline order is absent for $d\leq 4$.
Recently, this type of behavior in the system with infinite persistence time has also been observed in Ref.~\cite{shi2024} and is related to the randomly quenched systems, as we will discuss below~\cite{Imry-Ma1975, Ikeda2024PRE}.

Finally, we discuss the behavior of the static structure factor defined by $S(\bm q) = \expval*{|\delta\hat\rho(\bm q)|^2}/N$, where $\delta\hat\rho(\bm q)$ is the Fourier transformed density fluctuations.
Figure~\ref{fig4}(d) shows $S(\bm q)$ on the $q_y$-axis for various $R$. 
$S(\bm q)$ shows the Bragg peaks at $q_y=|{\bm G}|=
4\pi/(\sqrt{3}\ell_0)$. 
The inset of \fig{fig4}(d) shows that 
the height of the Bragg peak, $S_* = S(\bm G)$, is proportional to
$N$. 
This is another direct evidence of the crystalline order~\cite{chaikin2000}.
Moreover, we find that $S_*$ depends on $R$ as $S_*\propto e^{-W_R}$ in the large $N$ limit, where $W_R$ is defined by \eq{16} below (see Sec.~S5 of the SM~\cite{SI}).
The vertical axis of the inset of \fig{fig4}(d) is rescaled by $e^{-W_R}$.
Note that for $R=10\sigma$, $S_*$ is proportional to $N$ only for large $N$.
This is natural since, at $R=10\sigma$, the long-range translational order appears from relatively large $r$ (see \fig{fig2}). 
Furthermore, in the low-$q$ regime, the structure factor behaves as $S(\bm q) \sim q^2$.
In other words, the system is hyperuniform with the exponent of $\alpha=2$, which is identical to that found in the fluid phase~\cite{Li2019,kuroda2023_chiral} and the cRO model~\cite{Galliano2023PRL, Hexner2017}.
Since the density field is directly related to the longitudinal displacement by the continuity equation; 
$\hat\rho(\bm q,\omega) = -iq\hat u_\parallel(\bm q,\omega)$, $S(\bm q)$ is written as $q^2 S_\RM{u}^\parallel(\bm q)$. 
Combining the fitting function for $S_\RM{u}^\parallel(\bm q)$ discussed above, we conclude that $S(\bm q)=(qR)^2/[2(1+(\xi_\parallel q)^4)]$. 
The solid lines in \fig{fig4}(d) are this function plotted using the same fitting parameter $\xi_\parallel$. 
The agreement is excellent. 
Note that for small $q$'s, $S(\bm q) \sim (qR)^2/2$ and that the prefactor of $q^2$ is determined only by $R$.
This is in contrast to the observation in the fluid phase, where the prefactor depends sensitively on $\varphi$~\cite{kuroda2023_chiral}. 
We also address that when $D\neq 0$, $S(\bm q)$ converges to a
constant as $q\rightarrow 0$~\cite{Li2019, kuroda2023_chiral}, and
$S^\RM{\parallel,\perp}_\RM{u}(\bm q) \propto 1/q^2$, implying  the absence of hyperuniformity and translational order~(see Sec.~S6 of the SM~\cite{SI}).

{\it Linear elastic theory.}--- 
To understand the numerical observations theoretically, we develop a linear elastic theory for the two-dimensional chiral active crystal. 
We assume that the coarse-grained displacement field $\bm u(\bm r,t)$ obeys the following equation~\cite{Huang2021PRE, Shi2023PRL, Henkes2020Nature_Communications, chaikin2000}:
\begin{equation}
\p_t\bm u(\bm r,t) =\frac{1}{\zeta}\nabla\cdot \fdv{\mathcal F[\mathsf u(\cdot,t)]}{\mathsf u(\bm r,t)} + \bm \Xi(\bm r,t). \label{6}
\end{equation}
Here $\mathsf u(\bm r,t)$ stands for the strain tensor defined by 
$\mathsf u(\bm r,t) = (\nabla \bm u(\bm r,t) +\qty[ \nabla \bm u(\bm r,t)]^{\RM{T}})/2$.
The ``free energy'' functional $\mathcal F[\mathsf u]$ can be written as follows if the system is isotropic~\cite{landau1986,chaikin2000}:
\begin{equation}
\mathcal F[\mathsf u(\cdot)] = \frac{1}{2} \int_{V}\dd[2]\bm r\ \qty[\lambda \RM{Tr}[\mathsf u(\bm r)]^2 + 2\mu \mathsf u(\bm r):\mathsf u(\bm r) ],
\end{equation}
where $\lambda$ and $\mu$ are the Lam\'e coefficients.
$\bm \Xi(\bm r,t)$ denotes a Gaussian random field of zero mean.
In two-dimensional cABP, the random field has the following correlation~\cite{kuroda2023_chiral}:
\begin{align}
\expval{\bm\Xi(\bm r,t)\bm \Xi^\RM{T}(\bm r',t') } &= \frac{v_0^2\rho}{2}\mathsf R(\Omega(t-t'))\delta(\bm r-\bm r') \label{9} 
\end{align}
with
\begin{align}
\mathsf R(\theta)&=\mqty( \cos \theta & -\sin \theta\\ 
                                   \sin \theta&  \cos  \theta).
\end{align}
We now calculate the dynamical correlation function of the displacement defined by
\begin{equation}
S_\RM{u}^\RM{X}(\bm q,\omega) = \frac{1}{\rho}\int_{V}\dd[2]\bm r\int_{-\infty}^{\infty}\dd t\expval{{u}_\RM{X}(\bm r,t){u}_\RM{X}(\bm 0,0) }e^{-i(\bm q\cdot \bm r-\omega t)},
\end{equation}
where $\RM{X} \in\{\parallel,\perp\}$ and $\omega$ is the frequency.
Using the relation $\expval{\hat u_\RM{X} (\bm q,\omega)\hat u_\RM{X} ^*(\bm q,\omega') }/N = 2\pi S_\RM{u}^\RM{X}(\bm q,\omega) \delta(\omega-\omega')$, we have
\begin{equation}
S_\RM{u}^\RM{X}(\bm q,\omega) = \pi S_\RM{u}^\RM{X,eq}(\bm q,\omega) \qty[\delta(\omega - \Omega) + \delta(\omega + \Omega)] \label{11}
\end{equation}
with
\begin{equation}
S_\RM{u}^\RM{X,eq}(\bm q,\omega)  = \frac{1}{2} \frac{v_0^2}{\omega^2 + a_\RM{X}^2q^4}. \label{12}
\end{equation}
Here $a_\parallel = (\lambda + 2\mu)/\zeta$ and $a_\perp = \mu/\zeta$.
Equations~(\ref{11}) and (\ref{12}) tell us why the crystalline order is stable in this system:
\eq{12} is the dynamical correlation function of the displacement in the overdamped equilibrium systems of the effective temperature $T=v_0^2\zeta/2$, and its dynamical modes are characterized by the poles $\omega =\pm a_\RM{X} q^2$.
This excitation destroys the order in two-dimensional equilibrium crystals~\cite{chaikin2000}.
However, in the chiral active crystal, such excitation vanishes due to the term $\delta(\omega\pm\Omega)$, and the excited modes can exist only at $\omega = \pm \Omega$~\cite{Ikeda2024PRE}.
As a result, the fluctuations of the Goldstone mode, or the displacement field, are strongly suppressed and the order is stabilized~\cite{Ikeda2024PRE}.
This is distinct from other active crystals without chirality, where the activity contributes to additional excitations~\cite{Huang2021PRE, Shi2023PRL, Caprini2023JChemPhys, Caprini2023PRE, dey2024enhanced}.

The equal time correlation function of the displacement is calculated by integrating \eq{11} over $\omega$:
\begin{equation}
S_\RM{u}^{\RM{X}}(\bm q) = \frac{1}{2\pi} \int_{-\infty}^{\infty}\dd\omega\ S_\RM{u}^{\RM{X}}(\bm q,\omega) 
=\frac{1}{2} \frac{R^2}{1+(\xi_\RM{X}q)^4} \label{13}
\end{equation}
with the two correlation lengths: $\xi_\parallel = \sqrt{ (\lambda + 2\mu)/(\zeta\Omega)}$ and $\xi_\perp = \sqrt{ \mu/(\zeta\Omega)}$. 
Equation~(\ref{13}) explains the numerical results shown in \figs{fig4}(a) and (b).
Furthermore, our theory leads to $\xi_\parallel > \xi_\perp$, and $\xi_{\parallel,\perp} \propto \sqrt{R}$ since $\Omega \propto 1/R$. 
This is again in agreement with the numerical results~[see \fig{fig4}(c)].
From the above results, one can derive the asymptotic behavior of $C_{\bm G}(r)$ for large $r$ (see Sec.~S4 of the SM~\cite{SI} for the derivation):
\begin{align}
C_{\bm G}(r)
&= \exp[-\frac{ G^2}{8\pi\rho}\int_0^{\Lambda_\RM{c}}\dd q\ q\qty[1-J_0(qr)]S_\RM{u}(\bm q)] \notag \\
&\sim e^{-W_R} + O(r^{-1/2})\ \ \ (r\rightarrow \infty) \label{15}
\end{align}
with
\begin{equation}
W_R=\frac{ G^2R^2}{16\pi\rho}\qty[\frac{1}{\xi_\parallel^2}\arctan(\Lambda_\RM{c}^2\xi_\parallel^2) 
+\frac{1}{\xi_\perp^2}\arctan(\Lambda_\RM{c}^2\xi_\perp^2)], \label{16}
\end{equation}
where $J_0(z)$ is the Bessel function of the first kind. 
Equation~(\ref{15}) means that the long-range translational order exists as long as $R$ is finite. 
In \fig{fig3}(b), the solid lines and star symbols represent the first and second equations of \eq{15}, respectively.
Here we set $\Lambda_\RM{c} = 1/\ell_0$.
Equation~(\ref{15}) also explains $C_{\bm G}(r) - C_{\bm G}(\infty) \propto r^{-1/2}$ for large $r$ shown in the inset of \fig{fig3}(b).
Furthermore, from \eq{15}, one obtains $S_* \simeq  Ne^{-W_R}$, which supports our numerical observation (see the inset of \fig{4}(d) and Sec.~S5 of the SM~\cite{SI}).
Finally, we argue that for $\Omega =0$ (or $R\rightarrow\infty$), $\bm \Xi(\bm r,t)$ in \eq{9} does not depend on time and thus behaves like a quenched random field.
Indeed, \eq{13} in this limit leads to $S_\RM{u}^\RM{X}(\bm q)\propto 1/q^4$, which means that the order is unstable for $d\leq 4$. 
This is nothing but the Imry--Ma scenario for the quenched random systems with continuous symmetry where the lower critical dimension is four~\cite{Imry-Ma1975}. 
Hence, in two or three dimensions, the translational order parameter vanishes rapidly as the system size increases (see Sec.~S7 of the SM~\cite{SI}). 
We note that this scenario holds even in the presence of translational noise.

{\it Summary.}--- In this paper, we have numerically and theoretically studied the two-dimensional chiral and non-chiral active crystal using ABP with infinite persistence time ($D=0$).
When $\Omega$ is finite, the system develops true long-range translational order, and thus the crystals are very stable compared to equilibrium crystals. 
The crystalline phase is accompanied by hyperuniformity (HU)~\cite{Li2019,kuroda2023_chiral}. 
The suppression of the Goldstone mode, or the displacement field, caused by the same mechanism of HU is responsible for the emergence of long-range translational order. 
Considering the recent development of experiments in active matter~\cite{Kawaguchi:2017aa, Liu2019PRL, Li2019PNAS, Soni:2019aa, Deblais2020PRL, Zhang:2021aa, Shimaya2022, Iwasawa2021PRR}, especially the observations of HU in two-dimensional chiral active fluids~\cite{Huang2021, Zhang2022}, it may be possible to realize the two-dimensional chiral active crystal and experimentally confirm the true long-range translational order.   
We have also shown that the system has no crystalline order at $\Omega = 0$ because the system can be regarded as a randomly quenched system with continuous symmetry~\cite{Imry-Ma1975, Ikeda2024PRE}. 
As a future direction, it would be important to extend the present work to disordered solids~\cite{Mandal:2020aa, Szamel2024SoftMatter}. 
Since the Mermin--Wagner fluctuations are also present in the passive disordered solids, or glasses~\cite{Shiba2016PRL, Illing2017PNAS}, we speculate that similar stabilization or destabilization will be observed for active glasses.
A recent numerical study suggests the enhancement of the displacement fluctuations in two-dimensional disordered active solids with finite persistence time~\cite{dey2024enhanced}. 
Further numerical studies are needed to better understand this observation with the result of the present study.


\begin{acknowledgments}
We thank Grzegorz Szamel, Harukuni Ikeda, Kyosuke Adachi, and Hiroyoshi Nakano for fruitful discussions.
This work was supported by KAKENHI (Grant Number JP20H00128, JP22H04472, JP23H04503, JP23KJ1068, JP24H00192) and JST FOREST Program (Grant Number JPMJFR212T).
\end{acknowledgments}

%
\ifarXiv
    \foreach \x in {1,...,\numbersupplementpages}
    {
        \clearpage
        \includepdf[pages={\x,{}}]{\supplementfilename}
    }
\fi
\end{document}